\documentclass{ifacconf}

\usepackage{graphicx}      
\usepackage{natbib}        
\usepackage{epsfig} 
\usepackage{mathptmx} 
\usepackage{times} 
\usepackage{amsmath} 
\usepackage{amssymb}  
\usepackage{enumerate}
\usepackage{enumitem}
\usepackage{comment}
\usepackage{soul}
\usepackage{url}
\usepackage{caption}

\usepackage[dvipsnames]{xcolor}
\DeclareMathOperator*{\argmin}{arg\, min}
\DeclareMathAlphabet{\mathcal}{OMS}{cmsy}{m}{n}

\usepackage{multirow}
\begin{document}
\begin{frontmatter}

\title{Model-Based Control of Water Treatment with Pumped Water Storage} 

\author{Ryan Mauery,} 
\author{Margaret Busse,}
\author{Ilya Kovalenko} 

\address{Department of Mechanical Engineering, 
        The Pennsylvania State University, State College, PA 16801 USA}

\begin{abstract}
    Water treatment facilities are critical infrastructure they must accommodate dynamic demand patterns without system disruption. These patterns can be scheduled, such as daily residential irrigation, or unexpected, such as demand spikes from withdrawals for fire management.
The critical necessity of clean, safe, and reliable water requires water treatment control strategies that are insensitive to disturbances to guarantee that demand will be met.
One essential problem in achieving this is the minimization of energy costs in the process of meeting water demand, especially as the need for decarbonization persists.
This work develops a control-oriented hydraulic model of a water treatment facility with integrated pumped storage and introduces a model predictive control strategy for scheduling treatment plant system operations to minimize greenhouse gas emissions and safely meet water demand. 
\end{abstract}


\end{frontmatter}
\section{Introduction}

Access to clean water is essential to everyday life in today's society.
Therefore, safe and effective water treatment processes are considered critical infrastructure in the modern world \citep{textbook}.
However, water treatment systems can experience external disturbances such as drought, power outages, or contamination due to flooding \citep{korean_disasters}, and these effects will only be exacerbated over time by climate change.
The demand patterns that a treatment system must accommodate will also see changes, such as high volume flow for firefighting \citep{kanta_multiobjective_2012} or increased consumption rates due to COVID-19 lock downs \citep{abu-bakar_quantifying_2021}.
Therefore, the necessity of providing reliable, safe water requires modern water treatment to be robust to a wide range of disturbances. 

To efficiently and reliably adapt to disturbances while maintaining water quality, modern systems require advanced process control.
Advanced process control requires models of the water treatment process.
However, control-oriented models have not been designed for water treatment.
Through the use of control-oriented models, we can leverage advanced process control tools to effectively control water treatment.
For example, Model Predictive Control (MPC) has been used to compensate for system disturbances through a continuously simulated model in domains such as chemical processing \citep{mpc_chems} and hydraulic systems \citep{pangborn}.

Prior works have developed predictive control schemes that optimally control water distribution for reference tracking of flows (Georges, 1994), minimizing network pumping costs while meeting demand \citep{wang_non-linear_2017}, and employing pumped water storage to reduce distribution supply deficit \citep{sankar_optimal_2015,kurian_optimal_2018}.
However, these works have focused on ensuring demand-side optimization and  not on effective supply-side control.
Furthermore, the relevant literature lacks analysis for diverse scenarios or in-depth exploration of control schemes that address multi-objective systems \citep{castelletti_model_2023}. 
The crucial nature of a water treatment plant within our infrastructure network is more complex than a single node within the network and thus requires more robust, redundant, safe controls to ensure that set points are met. 
Currently, there is a lack of suitable models for optimal supply-side water treatment control that can be translated across a wide range of scenarios.
Therefore, there is a need for control-oriented water treatment models that can be used for flexible control strategies.

This work focuses on applying MPC to optimize water production rather than distribution, ensuring that a water treatment plant can treat water to an acceptable water quality standard (e.g., a chemical concentration during the treatment process), minimize greenhouse gas production, and output water as a constant-pressure, variable-flow source.
To address these objectives, this paper makes the following contributions (1) develops a dynamic system model for a water treatment plant with pumped water storage, (2) maps relevant water quality standards and energy scenarios to MPC controller objectives and constraints, and (3) evaluates the performance of the controller through a simulation case study.
The results present a functioning simulated model, and showcase that a predictive, model-based control algorithm can decrease pressure fluctuation and greenhouse gas production compared to a reactive controller.

The contents of this paper are organized as follows: Section 2 reviews and discusses relevant prior work. Section 3 introduces the water treatment process. Section 4 develops a mathematical model for simulating hydraulic systems. Section 5 introduces the MPC controller. Section 6 presents the case study, and Section 7 states the conclusions of the work.

\section{Background}

Water treatment plants must provide a water supply that can meet the time-varying flow consumed by its distribution network (\cite{EPANET}, 2020). 
This demand for water treatment can be modeled through deterministic explanatory variables derived from historical data, as well as stochastic variances due to unpredictable events \citep{kindler}.
These demand predictors are then used to design treatment plants that meet daily and seasonal demand cycles and address unexpected demand spikes.
The industry standard solution for minimizing energy costs due to fluctuating energy prices is to size storage tanks for clean water that fill during low demand, low-cost periods, and discharge during peak demand or low energy availability (\cite{AWWA}, 2013).
Optimization of integrated water and energy systems can be achieved through incorporation of operational flexibility by deliberately altering production patterns according to changes in energy supply, i.e., demand response \citep{kirchem_modelling_2020}.
Therefore, it is desired to take advantage of modern modeling and controls methods to optimize water treatment scheduling for reliability and efficiency.

\subsection{Hydraulic Modeling}
An uncontrolled water distribution network can be described as a linear system. 
Differential equations govern the evolution of the flows and pressures in reactive system elements, and algebraic equations constrain flow and pressure continuity throughout the system. 
Prior works have developed state space models around these equations to simulate system dynamics in discrete-time simulations \citep{georges_decentralized_1994,sankar_optimal_2015}.

Pumping and valve-switching actions to control water distribution are typically nonlinear. 
In prior works, these control actions are often simplified from continuous settings to boolean values:
Tank ports and other network valves are either switched open or closed, which allows control of the system to be reduced to a form that can be solved with Linear Programming methods \citep{kurian_optimal_2018, wang_non-linear_2017}.

\begin{figure}[tb]
    \centering
\includegraphics[width=1\columnwidth]{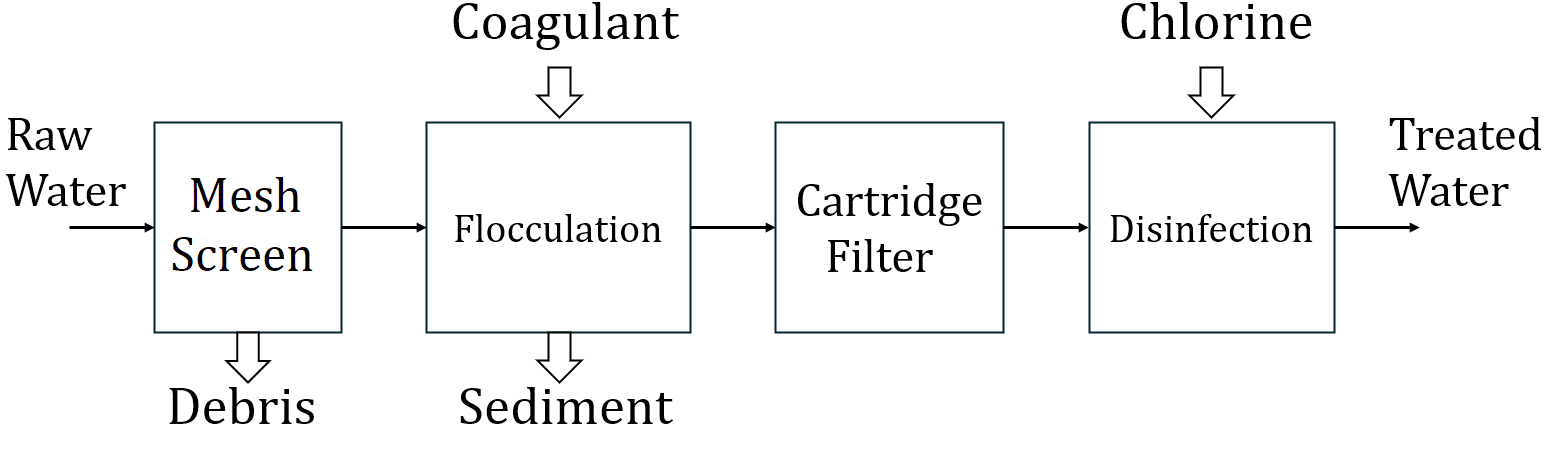}
    \vspace{-0.3cm}
    \caption{Simplified water treatment process flow chart}
    \label{fig:flow chart}
\end{figure}

In water treatment and distribution practice, the main energy consumption is associated with the pumps driving the system. 
Works that minimize flow or pressure deficiencies over a municipal water network consider pumps to fixed-speed and operate as a bang-bang actuator \citep{kurian_optimal_2018, wang_non-linear_2017}. This simplifies the pumping actions similarly to the discrete valve assumption, but decreases the flexibility of control over the system.
Other works take a more detailed approach by assuming variable-speed pumps, which allows for direct control of the generated pressure head \citep{sankar_optimal_2015}. 
While a variable-speed pump has better control over system pressure, this work does not consider the energy-saving potential of variable-speed pumping.

\subsection{Hydraulic Control}

Various control methods have been proposed in the literature to improve water distribution and treatment. 
In water distribution control, some works design their control objective function to optimize only for matching distribution with demand \citep{sankar_optimal_2015}. 
This minimizes pressure or flow shortages, but does not consider energy costs.
Other works minimize total energy consumption in addition to deviation from flow set points \citep{kurian_optimal_2018,wang_non-linear_2017}, but assume energy costs are not time-varying.

In water treatment control, studies have considered control objectives to satisfy consumer demand and control storage tank levels under plant constraints for a single input-output model using Linear-Quadratic Gaussian (LQG) adaptive control methods \citep{mourad_adaptive_2001}.
Furthermore, an urban water treatment plant was parametrically characterized for proposed MPC control, which was evaluated in a benchmark simulation that demonstrated flow reference tracking \citep{threetank}. 
However, this benchmark simulation was not performed in the context or scale of a municipal water treatment plant.
MPC methods have been shown to effectively schedule cost-optimized pumping and track reference trajectories for network distribution \citep{sankar_optimal_2015, kurian_optimal_2018, wang_non-linear_2017}.
However, existing works lack model and control of the water treatment system prior to distribution.

In addition, although application of modern control to water treatment has shown flow and pressure requirements being met, relevant phenomena such as time-varying energy pricing, greenhouse gas emissions costs, and treated water quality dynamics have not been addressed in water treatment control literature.
This paper will contribute by proposing a control-oriented model and MPC controller that considers greenhouse gas (GHG) emissions and the chemical kinetics of chlorine in water treatment and storage.

\section{Process Modeling}


We define flows entering the system as \textit{influent} and those exiting as \textit{effluent}. 
Influent water sources have varying water quality due to impurities
, so there are a broad range of treatment processes to consider \citep{textbook}. 
Plants are therefore designed to compensate for specific influent water quality parameters. 
Fig.~\ref{fig:flow chart} depicts a simplified process flow chart for a representative water treatment plant, derived from a US Environmental Protection Agency (EPA) survey of the processes represented in 8,073 US treatment plants \citep{epa_data}.

As shown in Fig.~\ref{fig:flow chart}, several steps have been identified that provide a high-level representation of a water treatment facility.
We assume that first, centimeter-scale debris suspended in the influent is removed before entering the treatment plant by the mesh screen.
This assumption is made so the flow can be simplified as a pure fluid, instead of a multiphase flow with discrete solid particles suspended.

\begin{figure}[tb]
    \centering
\includegraphics[width=1\columnwidth]{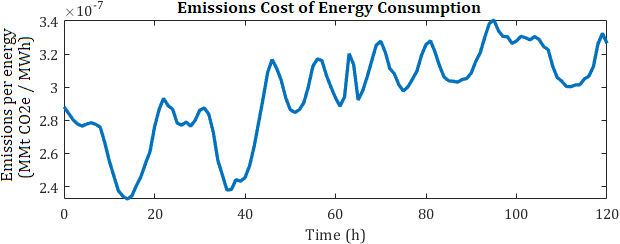}
\vspace{-0.4cm}
    \caption{Estimated $\phi(t)$, hourly CO2 emissions per unit energy consumed}
    \label{fig:energies}
\end{figure}

The next treatment process is coagulation, flocculation, and sedimentation. 
In this process, the water is chemically treated by rapid mixing of a coagulant such as alum \citep{textbook}.
The flow then moves to a sedimentation chamber, where dissolved particles on the millimeter scale will \textit{flocculate}, or aggregate into larger particles and settle in a sludge basin over a detention period.
Standard values for this detention period range from 1 to 30 minutes, which varies on the particulate matter chemical makeup and the mechanical design of the coagulant mixing and sedimentation chamber  \citep{textbook}. 
As described in Section 4, we propose to model flocculation process as a time delay on the order of minutes.

After coagulation and sedimentation, water is treated with a second filtration step that uses a cartridge filter.
This removes smaller particles down to the micron scale, including pathogens such as \textit{Giardia} \citep{textbook}. 
The EPA treatment plant survey indicated that cartridge filtration is the most common method for filtration of micron-scale particles \citep{epa_data}. 
Rapid cartridge filtration captures small particles at the expense of flow resistance, therefore inducing head loss as a function of filter porosity and filtration rate \citep{textbook}.
Note that since cartridge filters can be removed for cleaning \citep{textbook}, a dirty cartridge can be swapped with a clean one instead of backwashed, therefore backwashing will not be modeled. 
As described in Section 4, we assume a cartridge filter behaves as a simple resistance, where pressure loss is linearly proportional to the treatment flow rate.

The final step in a standard treatment process is disinfection to eliminate pathogens and prevent microbial growth in the distribution network. This can be achieved through chemical addition such as chlorination, energy-driven methods such as UV sterilization, or combined processes such as ozonation \citep{textbook}. The previously mentioned EPA survey shows that the single most common method for drinking water disinfection is chlorination \citep{epa_data}. 
Since free chlorine is reactive in water \citep{chlorine_residuals}, chlorine concentrations will be dynamically modeled as in section 4.2.

\subsection{Energy and Emissions Modeling}

The emissions generated from a system will depend on the energy mix, the proportion of active energy sources.
During times when renewable energy sources, such as wind and solar, are active, the emissions cost will be relatively lower. 
Contrarily, when renewable sources are inactive and auxiliary fossil fuel generators are activated to meet the energy demand, emissions cost per unit energy increases \citep{gagnon_cambium_2022}.

We model the time-varying emissions costs of energy by  calculating an hourly estimate for emissions per unit energy.
Publicly available data obtained from the National Renewable Energy Laboratory reports energy mixes and greenhouse gas generation for 2022 across 134 power grid networks in 48 US states \citep{gagnon_cambium_2022}.
A sample hourly emissions curve was determined by performing a least squares regression on the data to predict greenhouse gas generation as a linear combination of wind, solar, hydroelectric, gas, coal, and nuclear energy production.
The predictor coefficients are then multiplied by sample real-time energy mix data for a 5-day time period (\cite{eia_data}, 2024), selected to represent a Monday through Friday workweek.
The time series in Figure~\ref{fig:energies} depicts the resultant  emissions cost of energy time series used in this model.
The daily peaks and valleys occur because the low-carbon energy generated by wind and solar depends on time of day and weather conditions.

\section{Dynamic Modeling}
\label{sec:model}

The model is a non-linear discrete-time system with states \(x(k)\), control inputs \(u(k)\), and outputs \(y(k)\) in the following form
\begin{equation}
x(k+1)=g(x(k),u(k))
\end{equation}
\begin{equation}
y(k)=h(x(k),u(k))
\end{equation}
where the dimension of \(x\) is the number of dynamic elements, i.e. tanks, in the system, the dimension of \(u\) is the number of valve and pump control inputs in the system, and the dimension of \(y\) is the number of system outputs.
Link and node analysis of a network is used to determine the non-linear dynamic equations that will satisfy flow continuity and pressure cohesion for the model.
The flow through any pipe or the pressure at any node will be a function of the tank pressures \(x\) and the valve and pump settings \(u\), determined by the topology of the system.

\subsection{Passive Elements}
The two passive elements considered in modeling the processes are pipes and tanks. 
A pipe is modeled as an ideal dissipative element: pressure drop is linearly proportional to flow. 
Thus, the pressure difference between two nodes \(i\) and \(j\) is modeled by $    P_{j}(k)-P_{i}(k)=R_{ij}F_{i,j}(k)
$, 
where \(P_{j}(k)\) and \(P_{i}(k)\) are the high pressure ``upstream'' and the low pressure ``downstream'' node pressures at time step \(k\), \(R_{ij}\) is the pipe resistance between the nodes in units of pressure per volume flow,  and \(F_{i,j}(k)\) is the volume flow at time step \(k\) through the pipe from \(i\) to \(j\).

A tank is modeled as an ideal capacitive element. 
Assuming a constant tank cross section, the discrete time representation of tank pressure \(x_j\) at time step \(k+1\) in terms of the pressures and flows at time step \(k\) as follows:
\begin{equation}
    x_j(k+1)=x_j(k)+ {C_j}\Delta t \sum_{}^{i} F_{i,j}(k) 
\end{equation}
where \(C_j\) is the tank capacitance (pressure per volume), \(\Delta t\) is the time between steps \(k\) and \(k+1\), and \(F_{i,j}(k)\) and are the flows from nodes \(i\) into tank \(j\) at step \(k\).
Negative values of \(F_{i,j}(k)\) imply flow in the reverse direction.

Chlorine reacts with dissolved ionic species or metal structures such as tank walls and pipe surfaces.
Its decay is modeled at a bulk rate \(K\) proportional to the concentration \citep{chlorine_residuals}.
The concentration of chlorine in a tank \(y_c\) is modeled by 
\begin{equation}
     \frac{dy_C}{dt} = \sum{F_{i}c_{i}}-Ky_C 
\end{equation}
 where \(F_{i}\) are the flow rates into the tank, and \(c_{i}\) are the chlorine concentrations of the flows.
This model will simulate the chlorine concentration in effluent water storage after treatment.

\subsection{Active Elements}
The two active elements considered in modeling the process are valves and pumps. A valve is modeled as an ideal dissipative element whose resistance is actuated by a control input. Therefore, the flow between two nodes connected by a valve is described by the following nonlinear algebraic equation: $F_{i,j}(k)=r_ij(k)\cdot \left( P_{i}(k)-P_j(k)\right)$, where \(F_{i,j}(k)\) is the flow from a high pressure node \({i}\) to low pressure node \(j\) at time step \(k\), and \(r_{ij}(k)\) is the control input to the valve \(ij\). 
Since the flow across a valve is pressure driven, a nonzero control effort does not add energy to the system. 
It is assumed that the energy required to actuate the valves is negligible compared to pumping costs.

\begin{figure}[tb]
    \centering
    \includegraphics[width=1\columnwidth]{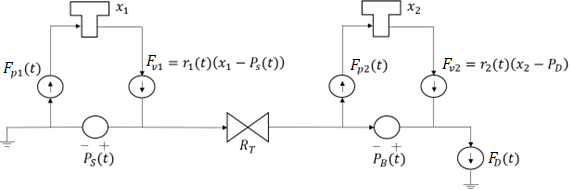}
    \vspace{-0.6cm}
    \caption{Flow chart for example hydraulic water treatment system}
    \label{fig:setup}
\end{figure}

Pumps are modeled as ideal sources that either prescribe a developed pressure difference between two nodes or maintain a prescribed flow rate through a pipe.
It is assumed that a lower level reactive (e.g., a PID controller) handles internal pump dynamics with a response time significantly faster than the time step length of the hydraulic system simulation.
The power consumed by an ideal pump, \(\mathcal{P}_{ij}(k)\), is equal to the work done on the pumped fluid, described by the following nonlinear algebraic equation:
\begin{equation}
    \mathcal{P}_{ij}(k)=F_{ij}(k) \cdot \left( P_{j}(k) - P_i(k)\right)
\end{equation}
The emissions rate for pumping at step \(k\) is therefore the product of emissions cost of energy and total power consumption: 
\begin{equation}
y_{E}(k)=\phi(k)\cdot\sum{\mathcal{P}_{ij}(k)}
\end{equation}
Component diagrams for the discussed active and passive system elements are hosted in a Github repository\footnote{https://github.com/RyanMauery/Water-Treatment-Modeling-MECC24}.
\subsection{Example system}
The steps for developing of the control-oriented model are to determine the system states, derive equations for the time-evolution of those states, and then represent the response variables in the output vector \(y(k)\) as a function of the system states and control inputs \(x(k)\) and \(u(k)\). 
Since tank pressures are governed by dynamic equations, each tank pressure in the system should be selected as a state.
To express the evolution of the tank pressure using states and the inputs, flow continuity at each node \(j\) to its surrounding \(i\) nodes is applied: $\sum_{}^{i}F_{i,j}(k)=0$ to generate one algebraic equation for each node.
The system is solved by substituting the characteristic equations for each element according to their respective link flows. 

Fig. ~\ref{fig:setup} depicts an example treatment plant flow diagram. 
The main pathway of the treatment process starts at a low pressure reservoir and flows through the pressure-controlled source pump $P_S$.
The source pump raises the node pressure to the source pump pressure, the treatment block $R_T$ induces time delay and pressure drop for flocculation and cartridge filtration, and the compensating booster pump $P_B$ regulates the effluent pressure.
Parallel to the source and pressure pumps are the raw and processed storage tanks $x_1$ and $x_2$. 
The flow-controlled pumps $F_{p1}$ and $F_{p2}$ connect to the tank inlets, and the tank outlets flow through the valves $F_{v1}$ and $F_{v2}$ respectively.

For this example, the distribution pressure is as follows:
\begin{equation}
\begin{split}
    y_D(k)=\frac{1}{1+r_2(k)R}\left[P_s+P_b(k)-R\left(F_D+F_{p2}(k)-r_2(k)x_2(k)\right) \right]
\end{split}
\end{equation}
This equation describes how the effluent pressure to the water distribution network correlates with tank states and input flows and pressures for the provided example. 
While this effluent pressure is specific to the example, it is possible to derive this pressure by analyzing the network of the target system. 
Using this information, we are able to stabilize the treatment plant using a predictive controller, as described in Section 5.

\section{Model Predictive Control Design}

Let the objective function \(J(k):U\subset \mathbb{R}^{m} \rightarrow \mathbb{R}\) map every point in the control space to a scalar cost at time step \(k\). 
The cost function is the quadratic sum of each scaled output cost, \(\lambda y_i^2\), summed over the prediction horizon \(N\).
For set point tracking instead of stabilization to zero, an output \(y\) is adjusted to \(\tilde{y}\) as follows $\tilde{y}(k)=y_{sp}-y(k)$, 
with \(y_{sp}\) the desired set point of the output.
This is used for the effluent pressure and chlorine concentration tracking.
Optimal control signals for time step \(k+1\) are those that minimize the cost function.
\begin{subequations}
\vspace{-0.1cm}
\label{eq:optimization}
\begin{align}
    u^{*}(k+1) &=\argmin_{u(k)\in U} \sum_{j=0}^{N-1} \lambda_C \tilde{y}_C^2 +\lambda_D \tilde{y}_D^2 +\lambda_E y_E^2
    \label{eq:objective}\\
    \operatorname*{s.t.} \quad 
    & x(k+1)=g(x(k),u(k)) 
    \label{eq:const1} \\    
    & y(k)=h(x(k),u(k)) 
    \label{eq:const4} \\
    & \underline{y_P} \leq y_P \leq \overline{y_P}, \quad \underline{x} \leq x \leq \overline{x}
\label{eq:const2}
\end{align}
\vspace{-0.52cm}
\end{subequations}

Where \(\tilde{y}_C, \tilde{y}_D, y_E, y_P\) are the chlorine concentration tracking error, distribution pressure tracking error, emissions generation, and pipe pressure at the \(k+j\) time step. 
Note that constraints \ref{eq:const1} and \ref{eq:const4} correspond to the dynamic model described in Section~\ref{sec:model}.
The limits \(\underline{y_P}, \overline{y_P}\) and \(\underline{x}, \overline{x}\) are the respective lower and upper bounds of pipe and tank pressures.
The cost weighting factor \(\lambda\) associated with each output normalizes the minimum output cost to unity and scales the output relative to its priority in the desired output performance.
For the case study, these values were selected by trial and error.
Future work will apply adaptive methods to determine the weights for water treatment systems.

The control and prediction horizons cover the current and next time step and are both 5 minutes. 
A short prediction horizon was chosen for low computational complexity and fast turnaround to confirm the feasibility of the control scheme.

A discrete low-pass filter is applied to the control signal before sending it to the actuator, 
improving control smoothness by filtering out high-frequency switching that may damage equipment \citep{threetank}.

The control space is set so \(U\subset \mathbb{R}^{m} \), where \(m\) is the number of control inputs to the system. 
\(U\) is a mesh partitioned by the resolution and bounded by the upper and lower limits for each of its dimensions according to the respective control signal.
The MPC controller inputs are therefore constrained and will not consider solutions outside the control effort bounds.

\section{CASE STUDY}

\subsection{Case Study Setup}
The source code for modeling, simulation, and control of the case study is also hosted in the Github repository\footnotemark[1] with documentation to reproduce the results of the study.  

A dynamic simulation was run to evaluate the performance of the controller in minimizing costs and stabilizing set points. 
The length of the simulation is set to 120 hours to examine the periodic behavior of the system in response to cyclical demand.    
\begin{table}
  \caption{Case Study system parameters}
  \label{fig:parameters}
  \includegraphics[width=.95\columnwidth]{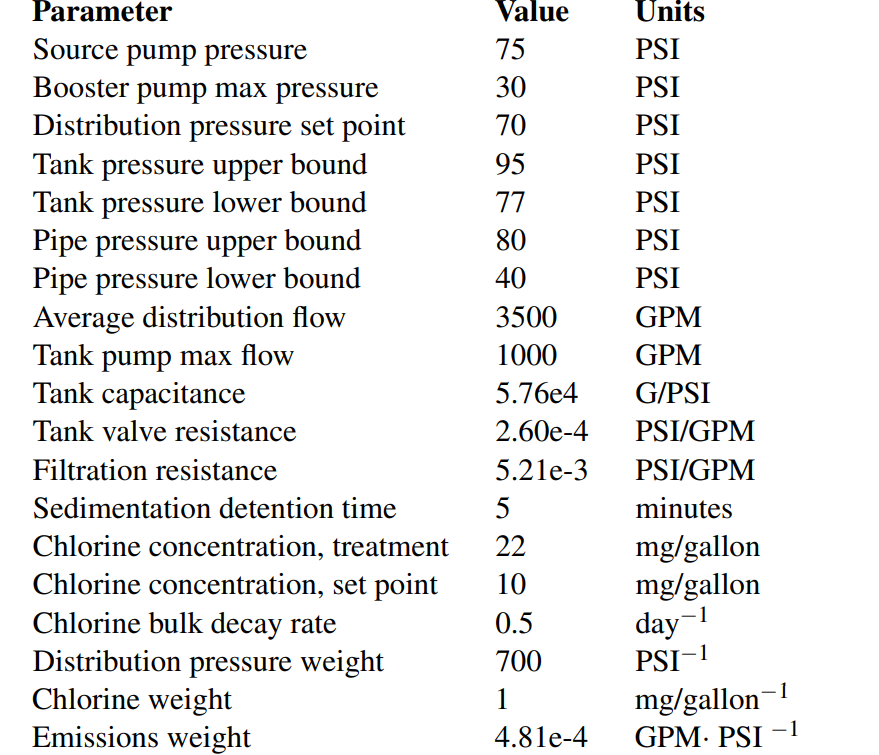}
\end{table}

The average throughput of the system stated in Table ~\ref{fig:parameters}, corresponds to a daily production of 5 million gallons. 
Appropriate component resistances and capacitances were determined from EPANET simulations (\cite{EPANET}, 2020). 
The pipe resistance was estimated by a linearization around the 3500 GPM operating point of a 12 inch diameter, 100 foot long pipe in the hydraulic solver. 
Tank capacitance is approximated from the tank time constant- the time elapsed for the tank to discharge by 63\%. In water treatment practice, levels of chlorine concentration are brought to 22 mg/gallon in the treatment plant by gaseous chlorine diffusion, and must be remain at no less than 6 mg/gallon before distribution \citep{textbook}. 
Literature chlorine values for the decay rate \(K\) range from 0.1 to 17 day\(^{-1}\) \citep{chlorine_residuals}, depending on water softness and tank chemistry.
The selected values for chlorine treatment are also reported in Table ~\ref{fig:parameters}.

A reactive controller designed heuristically is used as a benchmark.
The tank pump will run until the tanks are full, when it will then turn off the pumps and open the outlet valves until the tanks are empty.
The booster pump runs at a variable pressure, controlled by a reactive PI control law that attempts to stabilize the distribution pressure at its set point.
In existing industrial applications, this heuristic control law is implemented for simple tank cycling (\cite{EPANET}, 2020).


\subsection{Case Study Results}
Both schemes maintain safe levels of tank pressure in the simulation between the desired values of 77 and 95 PSI.
Since hard limits are set for reactive control, the system cyclically fills and empties both tanks, only considering the level of the tank when determining to turn on the inlet pumps or open the outlet valves. 
The reactive tank cycling frequency is dominated by the tank-time constant outlet valve setting. 
Since the valves are either full-open or full-closed, the flows through the outlet valves are approximately constant in the operating cycle pressures. 
On the other hand, the predictive tank levels appear to fluctuate irregularly.
When demand and emissions costs are high, the tanks are deliberately opened to take advantage of the previously-pumped stored water. 
When demand subsides and costs decrease, the valves close again and pumping restarts.

\begin{figure}[tb]
    \centering
    \includegraphics[width=1\columnwidth]{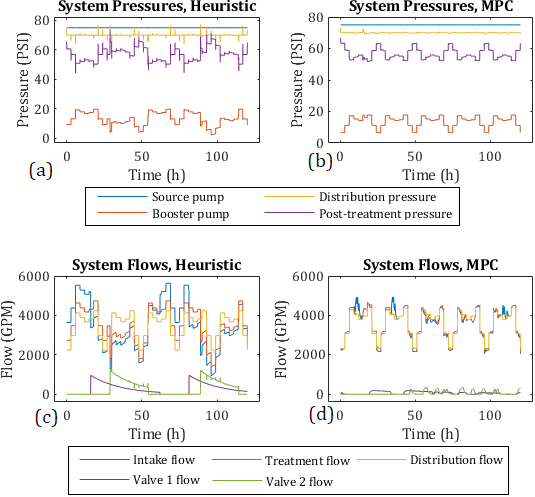}
    \caption{Comparison of system pressures and flows for heuristic (a and c) and MPC (b and d) controlled systems.}
    \label{fig: first system}
\end{figure}

\begin{figure}[tb]
    \centering
    \includegraphics[width=1\columnwidth]{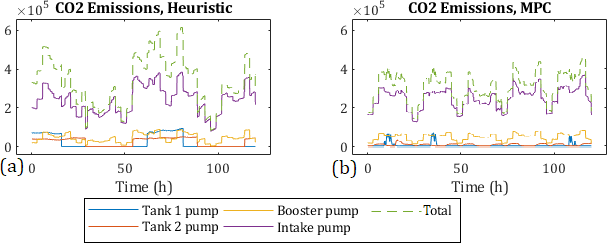}
    \caption{System emissions cost comparison}
    \label{fig: first emission}
\end{figure}

Since the reactive system tank levels are not synchronized with the demand cycle, the reactive booster pressure does not follow a daily cycle and has irregular peaks and valleys as shown in Fig. \ref{fig: first system}a. 
The predictive controller has a smoother distribution pressure in Fig. \ref{fig: first system}b.
Additionally, since the tank charging and discharging is smoother, the pressure at the system nodes will also be smoother.

Since the tank flows of the predictive system in Fig. \ref{fig: first system}d line up with the demand flow, system flow rates are generally less extreme than the reactive system in Fig. \ref{fig: first system}c.
In the reactive system, the higher maximum flow means  larger pipes will have poorer utilization during periods of lower flow.
In the reactive system, the intake and treatment flows regularly spike past or fall below the distribution flow.

\begin{table}
  \caption{Pressure and Emissions Costs Comparison}
  \label{tab:performance}
  \includegraphics[width=1\columnwidth]{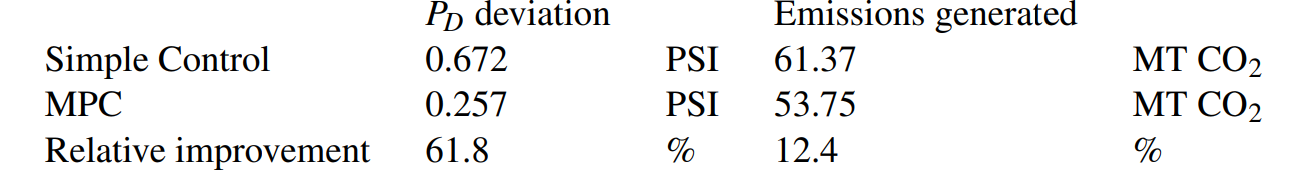}
\end{table}

Because the control inputs are selected to minimize short-term costs, the predictive control costs shown in Fig. \ref{fig: first emission}b again outperforms the reactive control costs shown in Fig. \ref{fig: first emission}a with a lower peak system emissions rate and a 12\% savings on total emissions production.

\begin{figure}[tb]
    \centering
    \includegraphics[width=1\columnwidth]{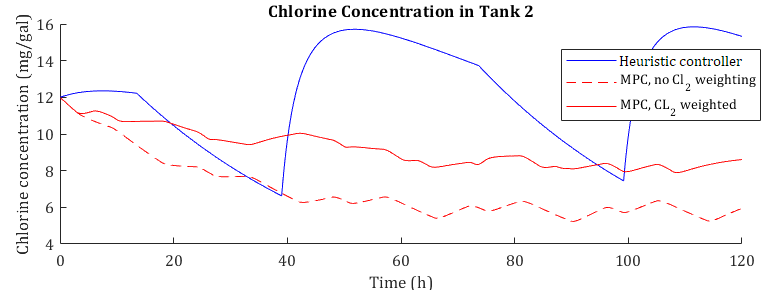}
    \caption{Chlorine concentration in Tank 2}
    \label{fig: chlorine}
\end{figure}

One of the results of the reactive control scheme is that the periodic turnover of tank 2 guarantees chlorine levels never fall too low, without requiring closed-loop feedback for water quality.
Fig. ~\ref{fig: chlorine} shows the chlorine concentration in tank 2 rising and falling as the tank fills and empties. 
For the MPC controller, if chlorine concentration is not considered, the clean water storage will be inadequately cycled and chlorine concentrations will decay below the safe minimum.
Implementing a set point and weight for chlorine concentration in the MPC cost function as detailed in Table ~\ref{fig:parameters} motivates the system to cycle freshly treated water into the storage tank.

\section{CONCLUSION}


In this paper, a control-oriented model was developed for a water treatment facility and a dynamic modeling framework is applied to a hydraulic system to evaluate the performance of MPC methods on meeting system objectives for a water treatment plant.
The controller has been simulated in a case study to show improvement for hydraulic performance as well as emissions and chemical objectives specific to sustainable water treatment.

This study is limited by the assumption that the model precisely predicts the behavior of the true system.
This is a result of making simplifying assumptions such as sufficiently fast low-level control, linear system element dynamics, and knowledge of true system parameters.
Further studies will be refined by incorporating complexities such as model mismatch, state uncertainty due to process noise, or dynamic state observation to compensate for poorly instrumented systems.

Ongoing work is investigating the robustness of the controller with further case study simulating disruptive scenarios such as pump fouling and load shedding.
A planned future extension to this work is development and experimental testing of a physical hydraulic testbed to test the proposed model and controller in more realistic scenarios.

\bibliography{references}

\end{document}